\documentstyle[aps,prb,multicol]{revtex}

\begin{document}
\title{Feasibility of biepitaxial YBa$_2$Cu$_3$O$_{7-x}$ Josephson junctions \\
for fundamental studies and potential circuit implementation}
\author{F. Tafuri}
\address{Dipartimento di Ingegneria dell'Informazione, Seconda Universit\`{a} di\\
Napoli, 81031 Aversa (CE) and\\
INFM-Dipartimento Scienze Fisiche dell'Universit\`{a} di Napoli ''Federico\\
II'', 80125 Napoli (ITALY)}
\author{F. Carillo, F. Lombardi, F. Miletto Granozio, F. Ricci, U. Scotti di Uccio
and A. Barone}
\address{INFM-Dipartimento Scienze Fisiche dell'Universit\`{a} di Napoli ''Federico\\
II'', 80125 Napoli (ITALY)}
\author{G. Testa and E. Sarnelli}
\address{Istituto di Cibernetica del CNR, Via Toiano 6, Arco Felice (NA) (ITALY) also%
\\
INFM}
\author{J.R. Kirtley}
\address{IBM T.J. Watson Research Center, P.O. Box 218, Yorktown Heights, NY 10598,\\
USA}
\date{\today}
\maketitle

\begin{abstract}
We present various concepts and experimental procedures to produce
biepitaxial YBa$_{2}$Cu$_{3}$O$_{7-x}$ grain boundary Josephson junctions.
The device properties have an interesting phenomenology, related in part to
the possible influence of ``$\pi $-loops''. The performance of our junctions
and Superconducting Quantum Interference Devices indicates significant
improvement in the biepitaxial technique. Further, we propose methods for
fabricating circuits in which ``0-'' and ``$\pi $-loops'' are controllably
located on the same chip.
\end{abstract}

\pacs{74.50.+r,74.80.Fp,85.25.Cp}

\preprint{}

\begin{multicols}{2}
\narrowtext

\pagebreak
\narrowtext
\section{Introduction}
The possibility of realizing electronic
circuits in which the phase differences of selected Josephson
junctions are biased by $\pi$ in equilibrium is quite
stimulating. \cite{ioffe}
The concept of such $\pi$-phase shifts was
originally developed in the ``extrinsic" case for
junctions with ferromagnetic barriers \cite{bulae} and in the
``intrinsic" case for junctions exploiting superconductors with
unconventional order parameter symmetries. \cite{larkin} As a result of
the possible d$_{x^2-y^2}$
order parameter symmetry of high critical temperature
superconductors (HTS), \cite{kirtley}
the presence of intrinsic $\pi$ loops
has also been considered for HTS
systems. \cite{sigrist} This has been
discussed recently in view of  novel device concepts, and in particular
for the implementation of a solid state qubit
\cite{ioffe,zago,mannh,ilic} and for Complementary Josephson junction
electronics. \cite{beas}
In this paper we
discuss how  YBa$_2$Cu$_3$O$_{7-x}$ (YBCO) structures made by the
biepitaxial technique \cite{tafuri,char} can be successfully employed
to produce arbitrary circuit geometries
in which both ``0" and $\pi$-loops are present, and possibly to obtain
a doubly degenerate state. \cite{ioffe,zago}
Of course, great caution should be used because of  stringent requirements on
junctions parameters for practical applications of such devices.

Josephson junctions
based on artificially controlled grain boundaries have been widely
employed for fundamental studies on the
nature of HTS. \cite{kirtley,mannh,ilic}  The lack of a reliable
technology based on  the traditional trilayer configuration (i.e. a sandwich type junction
with an insulator between the two superconducting electrodes) also
enhanced interest in GB Josephson junctions for
applications.  Although the mechanism of high-T$_C$
superconductivity and the influence of grain boundaries on the
transport properties are not completely determined, reproducible
and good quality devices are routinely fabricated. YBCO GB
junctions are usually classified as bicrystals, \cite{dimos}
biepitaxials, \cite{char} and  step-edges, \cite{simon} depending
on the fabrication procedure. The bicrystal technique typically
offers junctions with better performances and allows in principle
the realization of all different types of GBs ranging from [001] and
[100] tilt to [100] twist boundaries. GB junctions based on the
step-edge and biepitaxial techniques offer the advantage,
with respect to the bicrystal technology, of
placing the junctions on the substrate without imposing any
restrictions on the geometry. A comparison between the different GB
techniques is far beyond the aim of this paper. Nevertheless we
intend to show that significant improvements with respect to the
original technique developed by Char et al. \cite{char} are
possible for biepitaxial junctions, and that the resulting devices
have potential for applications. As a matter of fact, in
traditional biepitaxial junctions, the seed layer used to modify
the YBCO crystal orientation on part of the substrate produces
an artificial 45$^\circ$  [001] tilt (c-axis tilt) GB.
The nature of such a GB
seems to be an intrinsic limit for some real applications. A
convincing explanation has been given in terms of the d-wave
nature of the order parameter and more specifically by the
presence of $\pi$-loops. \cite{hilg} As demonstrated by studies on
bicrystals, based on the same type of 45$^\circ$  [001] tilt
GB, the presence of $\pi$ -loops reduces the I$_C$R$_N$ values (where
I$_C$ and R$_N$ are the critical current and the high normal state
resistance respectively), produces a dependence of the
critical current I$_C$ on the magnetic field H quite different from the
Fraunhofer-like pattern,  and generates
unquantized flux noise at the grain boundary. \cite{hilg}

We will
show that the implementation of the biepitaxial technique
\cite{tafuri} we developed to obtain 45$^\circ$  [100]  tilt and
twist (a-axis tilt and twist) GBs junctions makes such a technique interesting
for both applications and fundamental studies. The phenomenology observed for
the junctions based on these GBs and Scanning SQUID Microscopy investigations
demonstrate the absence of $\pi$-loops, as we expect from their microstructure.
As a consequence higher
values of the I$_C$R$_N$ values, a Fraunhofer like dependence of
I$_C$ on the magnetic field and lower values of the low
frequency  flux noise, when compared with 45$^\circ$  c-axis tilt
GBs, have been measured. These features are important tests to
employ junctions for applications. Scanning SQUID Microscopy
investigations also gave evidence
of ``fractional" vortices in the presence of impurities. Finally, we extended
the biepitaxial process to other types of GB  by
using different seed layers to obtain junction configurations
where $\pi$ loops can be controllably produced. We shall not dwell on
conceptual principles and actual feasibility of qubit devices. Instead we discuss
the importance of the biepitaxial technique in having ``0" and ``$\pi$" loops on the
same chip. This makes the biepitaxial technique more versatile
and promising for circuit design.

\section{Devices: concepts and fabrication procedure}
As mentioned above, the biepitaxial technique allows the fabrication
of various GBs by growing different seed layers and using substrates
with different orientations.  We have used
MgO, CeO$_2$ and SrTiO$_3$ as seed layers. The MgO and CeO$_2$
layers are deposited on (110) SrTiO$_3$ substrates, while
SrTiO$_3$ layers are deposited on (110) MgO substrates; in all these cases the seed
layers grow along the [110] direction. Ion
milling is used to define the required geometry of the seed layer
and of the YBCO thin film respectively, by means of  photoresist
masks. YBCO films, typically 120nm in thickness,  are deposited
by inverted cylindrical magnetron sputtering at a temperature of
780$^\circ$ C. YBCO grows along the [001] direction on MgO (substrates
or seed layers) and on the CeO$_2$  (seed layers), while it grows
along the [103]/[013] direction on SrTiO$_3$ (substrates or seed layers). In
order to select the [103] or [$\overline{1}$03] growth and to ensure a better
structural uniformity of the GB interface, we have also
successfully employed vicinal substrates. However, most of the
transport properties presented in this paper refer to samples not
using vicinal substrates. Detailed structural investigations
on these GBs, including Transmission Electron Microscopy (TEM)
analyses,  have been performed  and the results have been
presented elsewhere. \cite{tafuri,verb}

Depending on the
patterning of the seed layer and the YBCO thin film, different
types of GBs ranging from the two ideal limiting cases of
45$^\circ$  a-axis tilt and  45$^\circ$  a-axis twist have been
obtained (see Fig.1). The intermediate situation occurs when the
junction interface is tilted at an angle $\alpha$ different from 0 or
$\pi$/2 with respect to the a- or b-axis of the [001] YBCO thin
film. In all cases, the order parameter orientations do not
produce an additional $\pi$ phase shift along our junction, in
contrast with the 45$^\circ$  asymmetric [001] tilt  junctions. As a
consequence, no $\pi$ loops should occur independently of the details
of the interface orientation. In Fig. 1 we consider ideal
interfaces and  neglect meandering of the GBs or interface
anomalies that will be considered below. The CeO$_2$ seed layer
may produce a more complicated GB structure, in which a 45$^\circ$  c-axis
tilt accompanies the 45$^\circ$  a-axis tilt or twist (see
Fig.2a). \cite{scot} In this case, as shown in Fig.2b, $\pi$ loops should
occur in analogy with the traditional biepitaxial junctions based on
45$^\circ$  c-axis tilt GBs. In both Figs. 1 and 2 we display the possible
d$_{x^2-y^2}$ -wave order
parameter symmetry in the junction configuration. Junctions were
typically 4 microns wide. We also performed systematic
measurements on SQUIDs based on the structure employing MgO as a
seed layer and SrTiO$_3$ as a substrate. DC SQUIDs in different
configurations and with
loop inductance typically ranging from 10 to 100 pH have been investigated.
The typical loop size leading to the 10(100) pH inductance is approximately
10$^2$$\mu$m$^2$ (10$^4$$\mu$m$^2$).

\section{Experimental results}
\subsection{Biepitaxial junctions employing MgO seed layers}
In this
section we attempt to cover most of the phenomenology of the
transport properties of 45$^\circ$  a-axis tilt and twist
biepitaxial junctions. In Fig. 3,  current vs voltage (I-V)
characteristics of a typical biepitaxial  junction are given for
various temperatures close to the critical temperature. In the
inset the corresponding I-V characteristic at T = 4.2 K is
reported. They are closely described by the
resistively-shunted-junction (RSJ) model and no excess current is
observed. Nominal critical current densities J$_C$ of  5 x
10$^2$A/cm$^2$ at T = 77 K, and of 9 x 10 $^3$A/cm$^2$ at T = 4.2 K
have been measured respectively. The R$_N$ value (3.2 $\Omega$)
is roughly independent of the temperature for $T<T_C$, providing
a normal state specific conductance $\sigma_N$ = 70
($\mu$$\Omega$cm$^2$)$^{-1}$. The maximum working temperature T$_C$
of this device was 82 K. In this case I$_C$R$_N$ is 1.3 mV at T =
4.2 K. These values typically ranged from 1 mV to 2 mV at T =
4.2 K. They are larger for the corresponding J$_C$ values than
those provided by conventional biepitaxials, and are of the same
order of magnitude as in GB bicrystal and step edge junctions.
\cite{tafuri} While the values of critical current density and
normal state specific conductance in the tilt case are quite
different from the twist case, the I$_C$R$_N$ values are
approximately the same for both. Moreover I$_C$R$_N$ does
not scale with the critical current density. \cite{tafuri} In the
tilt cases J$_C$ $\approx$ 0.5-10 10$^3$ A/cm$^2$ and
$\sigma$$_N$ $\approx$ 1-10 ($\mu$$\Omega$cm$^2$)$^{-1}$ are measured at T = 4.2
K respectively. Twist  GBs junctions are typically characterized
by higher values of J$_C$ in the range 0.1-4.0 x 10$^5$ A/cm$^2$ and
of $\sigma$$_N$ in the range 20-120 ($\mu$$\Omega$cm$^2$)$^{-1}$ (at
T = 4.2 K).  For the twist case deviations from  the RSJ model are more
marked as a result of higher critical current densities. For high values of  J$_C$ GB
junctions do not present any clear modulation of the critical current as a function
of the magnetic field.

A demonstration of the possibility of tailoring
the critical current density and of the different transport
regimes occurring in the tilt and twist cases has been given by
measuring the properties of junctions with different orientations
of the GB barrier on the same chip.
By patterning the seed layer as shown in Fig.4a, we could measure
the properties of a tilt junction and of junctions whose
interface is tilted in plane by an angle $\alpha$ = 30$^\circ$,
45$^\circ$  and 60$^\circ$  with respect to the a- or b-axis of the
[001] YBCO thin film respectively. In all cases the order
parameter orientations do not produce an additional $\pi$ phase shift
along our junction, in contrast with the 45$^\circ$ [001]  tilt
junctions, and no $\pi$ loops should occur. We
measured the expected increase of the critical current density with
increasing angle,
which corresponds on average to an increase of the twist current
component. The values measured at T = 4.2 K are reported in Fig.
4a and range from the minimum value J$_C$ =  3 x 10$^2$ A/cm$^2$
in the tilt case to the maximum  J$_C$ =  10$^4$ A/cm$^2$
corresponding to an angle of 60$^\circ$, for which the twist
component is higher. The consistency of this result has been
confirmed by the values of normal state resistances, which are
higher in the tilt case and decrease with increasing
$\alpha$. The I$_C$R$_N$ values are about the same for all
the junctions independently of the angle $\alpha$. In Fig. 4b the
I-V characteristics measured at T = 4.2 K, corresponding to the
junctions of Fig.4a, are shown for approximately the same voltage
range. Deviations from RSJ behavior appear for higher values
of the critical current density ($\alpha$ = 60$^\circ$). These
results demonstrate that the grain boundary acts as a tunable
barrier. This possibility of modifying the GB macroscopic
interface plane by controlling the orientation of the seed
layer's edge is somehow equivalent to the degree of freedom
offered by bicrystal technology to create symmetric or asymmetric
GBs, with the advantage of placing all the junctions on the same
substrate. The 45$^\circ$  a-axis tilt and twist GBs and the
intermediate situations can represent ideal structures to
investigate the junction physics in a wide range of
configurations.  The anisotropy of the (103) films and the
possibility to select the orientation of the junction interface
by suitably patterning the seed layer, and eventually the use of
other seed layers which produce different YBCO in plane
orientations, allow the fabrication of different types of junctions and
the investigation of different aspects of HTS junction phenomenology.
In particular we refer to the possibility of changing the tunneling
matrix elements (by selecting the angle $\alpha$) and to use the
anisotropy of the layered structure of YBCO properties and of the
order parameter symmetry.

The study of the junction properties
in the presence of an external magnetic field H is a fundamental
tool for the investigation of the Josephson effect in the various junctions,
as well as a test of junction quality. \cite{barone} We
observe  modulations of the critical current I$_C$ following the
usual Fraunhofer-like dependence. The   I$_C$ (H) patterns are
mostly symmetric around zero magnetic field, and in all samples
the absolute maximum of  I$_C$ occurs at H=0. The
presence of the current maximum at zero magnetic field is consistent with
the fact that in our junction configuration the order parameter
orientations do not produce an additional $\pi$ phase shift, in
contrast with the 45$^\circ$  [001] tilt GB junctions.
\cite{hilg,tafuri} Some examples are given in Fig. 5, where the
magnetic pattern relative to a SQUID and a single junction at T = 4.2 K
are shown respectively. In the former
case we can also distinguish a smaller field modulation
(with a period of 8 mG) which  corresponds to the SQUID
modulation (inset a).
In the latter case the I-V characteristics are reported for different
magnetic fields (inset b). Despite the Fraunhofer-like dependence,
some deviations
are evident, in agreement with most of the data available in literature.

For sake of completeness we also acknowledge some work we carried out by investigating 
Fiske steps as a function of H in other junctions, giving some evidence
of a dielectric-like behavior \cite{gross} of some of the layers
at the junction interface. We already reported about this work elsewhere. \cite{fran}
The Fiske steps do not
depend on the use of a particular substrate, since they have been
observed in junctions based both on SrTiO$_3$ and MgO substrates.
Typical values of the ratio between the barrier thickness t and
the relative dielectric constant $\epsilon_r$ range from 0.2
nm  to 0.7 nm. Considerations on the dependence of I$_C$ on the temperature (T)
can be also found in Ref.\cite{fran}. 
In junctions
characterized by lower critical current densities, I$_C$ tends to
saturate at low temperatures, in contrast to those
characterized by higher critical currents, for which there is a
linear increase.\cite{tafuri,bis}

\subsection{Scanning SQUID microscopy on  biepitaxial junctions with MgO seed layer}
Figure 6 is a scanning SQUID microscope\cite{ssmapl} image of a
200x200$\mu$m$^2$ area along a grain boundary separating a (100) region
from a (103) region (as labelled in the figure) of a thin YBCO biepitaxial
film grown as described above. The position of the grain boundary is
indicated by the dashed line. The image was taken at 4.2K in liquid helium
with an octagonal SQUID pickup loop 4 microns in diameter after cooling the
sample in a few tenths of a $\mu$T externally applied magnetic field normal
to the plane of the sample. The grey-scaling in the image corresponds to a
total variation of 0.13$\Phi_0$ of flux through the SQUID pickup loop.
Visible in this image are elongated interlayer Josephson vortices in the
(103) area to the right, and ``fractional" vortices in the (100) area to
the left, of the grain boundary. Fits to the interlayer vortices give a
value for the $c$-axis penetration depth of about 4$\mu$m. The
``fractional" vortices are spontaneously generated in the (100) film,
regardless of the value of external field applied.\cite{tafgbj} Temperature
dependent scanning SQUID microscope imaging shows that this spontaneous
magnetization, which appears to be associated with defects in the film,
arises when the film becomes superconducting.\cite{kirtley,bailey}
Although it is difficult to
assign precise values of total flux to the ``fractional" vortices, since
they are not well separated from each other, fits imply that they have less
than $\Phi_0$ of total flux in them, an indication of broken time-reversal
symmetry. Although there is apparently some flux generated in the grain
boundary region, the fact that these SQUIDs have relatively low noise seems
to indicate that this flux is well pinned at the temperatures at which the
noise measurements were made. These results are consistent with the absence
of $\pi$ loops along the grain boundary.

\subsection{ Biepitaxial junctions employing CeO$_2$ seed layers}
The CeO$_2$ seed layer, as anticipated in section II, may produce an
artificial GB that can be seen as a result of two rotations: a
45$^\circ$ [100] tilt or twist followed by a 45$^\circ$ [001] tilt
around the c-axis of the (001) film. For this junction
configuration a d-wave order parameter symmetry would produce
$\pi$-loops, as shown in Fig. 2. We notice that such $\pi$-loops are
structurally different from those usually obtained by the
45$^\circ$ [001] tilt GB junctions based on the  traditional
biepitaxial and bicrystal techniques. Due to the microstructure we expect
especially in the [100] tilt case low critical current densities and high normal state resistances.
We found that the deposition
conditions to select the uniform growth of YBCO 45$^\circ$ tilted
around the c-axis of the (001) film are critical.
Preliminary measurements realized on  tilt-type junctions with a CeO$_2$
seed layer gave evidence of Josephson coupling in these GBs.
The measured I$_C$R$_N$ values are from 200 $\mu$V to  750 $\mu$V and are in the
typical range of the GBs Josephson junctions.

\subsection{Biepitaxial SQUIDs employing MgO seed layers}
In this section
we report on the characterization of dc-SQUIDs which are to our
knowledge  the first employing the GBs discussed above.
\cite{testa} These SQUIDs exhibit very good properties, and
noise levels which are among the lowest ever reported for
biepitaxial junctions. \cite{testa} Apart from implications for
applications, these performances are important for the study of the transport
properties of HTS Josephson junctions. In Fig. 7 we show the
magnetic field dependence of the voltage at 77 K for different values of
the bias current for a dc-SQUID with an inductance of
13 pH. At this temperature I$_C$R$_N$ is about 20 $\mu$V. The
corresponding value of the screening parameter
$\beta$=2LI$_C$/$\Phi_O$ is 0.03. In
general low $\beta$ values are mandatory to avoid the influence of
asymmetric inductances in SQUID properties, and this has been crucial for
experiments designed to study the order parameter symmetry.\cite{mannh} The
presented curves are quite typical. These SQUIDs usually work in
a wide temperature range from low temperatures (4.2 K ) up
to temperatures above  77 K. The maximum working temperature was
in this case 82 K. The achieved magnetic flux-to-voltage transfer
functions V$_{\Phi}$ = $\partial$V/$\partial \Phi$, where V and
$\Phi$ are the voltage across the device and the applied magnetic
flux in the SQUID loop respectively, are suitable for
applications. For instance at T = 77 K an experimental value of
the SQUID amplitude voltage modulation $\Delta$V of 10.4 mV was
measured, corresponding to V$_{\Phi}$ = 36.9 $\mu$V/$\Phi_0$.
\cite{testa} Steps of different nature have been recurrently
observed in the I-V characteristics in the washer and hole
configurations and characterized also in terms of the magnetic
field dependence of the voltage at different values of the bias
current.

The noise spectral densities of the same
dc-SQUID have been measured at T = 4.2 K and T = 77 K using
standard flux-locked-loop modulated electronics. The energy
resolution $\epsilon$ = S$_{\Phi}$/2L (with S$_{\Phi}$ being the
magnetic-flux-noise spectral density) at T = 4.2 K and T = 77 K
is reported in Fig. 8. At T = 4.2 K and 10 kHz, a value of S$_\Phi$
= 3 $\mu \Phi_0$/$\sqrt{Hz}$ has been measured, corresponding to an energy
resolution $\epsilon$ = 1.6x10$^{-30}$ J/Hz. This value is the
lowest reported in the literature for YBCO biepitaxial SQUIDs.
Moreover, the low frequency 1/f flux noise spectral density at 1
Hz is more than one order of magnitude lower than the one
reported for traditional biepitaxials, as is also evident from
the comparison with data at T = 4.2 K of Ref. \cite{clarke}. The
lower values of low frequency noise are consistent with the
absence of $\pi$-loops on the scale of the faceting for these types
of GBs, as clearly shown by Scanning SQUID Microscopy results.
The $\pi$-loops produce some types of spontaneous magnetic flux
in the GB region, which among other
effects tends to degrade the SQUID's noise levels. \cite{hilg}

\section{Biepitaxial junctions for experiments on the symmetry of
the order parameter and for a development of concepts for  qubits}
The particular junction configurations investigated in this work
allow some consideration of the possible impact of these
types of junctions on the study of the Josephson effect and the order
parameter symmetry in YBCO and on the development of concepts for
devices. \cite{ioffe,zago,beas,mannh} We first recall that the
biepitaxial technique can provide circuits composed completely of
junctions without any $\pi$-loops (see Fig. 9a). By varying the
interface orientation with respect to the [103]  electrode
orientation, the junction properties can be adjusted. On the
other hand the traditional biepitaxial technique, \cite{char}
producing
45$^\circ$ [001] tilt GBs (see Fig. 9b) or the types of junctions
described in
the previous section by using CeO$_2$ (see Fig. 9c), can controllably generate
$\pi$-loops on macroscopic scales.
In these schemes we use a corner geometry with a 90$^\circ$
angle. This angle $\alpha$ can be obviously tuned to enhance the effects
related to the phase shift (see dashed line in Fig.9b) and this
change is particularly easy to realize by using the biepitaxial
technique.

In this section we focus our
attention mainly on the feasibility of the biepitaxial junctions
to obtain the doubly degenerate state required for a qubit.
In Ref.\cite{ioffe}$^a$ the design is based on quenching the lowest order 
coupling by arranging a junction with its normal aligned with the node
of the d-wave order parameter, thus producing a double periodic current-phase
relation. It has been shown that the
use of $\pi$ phase shifts in a superconducting phase qubit provides a
naturally bistable device and does not require external bias
currents and magnetic fields. \cite{ioffe}$^b$ The direct consequence
is the quietness of the device over other designs. A $\pi$
junction provides the required doubly degenerate fundamental
state, which also manifests itself in a doubly periodic
function of the critical current density as a function of the
phase. \cite{ilic} The same principle has been used in
small inductance five junction loop frustrated by a
$\pi$-phase shift. \cite{ioffe}$^b$
This design provides a perfectly degenerate two-level system and offers
some advantages in terms of fabrication ease and performance.
HTS may represent a natural solution for the realization of the required $\pi$-phase shift
due to the pairing symmetry of the order parameter and,
therefore, due to the possibility  of producing $\pi$
phase shifts.
Experimental evidence of YBCO $\pi$ -SQUIDs has been given by
employing the bicrystal technique on special tetracrystal
substrates. \cite{mannh} The biepitaxial technique, beyond
providing junctions with
opportune properties, would guarantee the versatility necessary for the
implementation of a real device, as shown below. As a matter of fact, we
notice that our technique
allows the realization of circuits where
$\pi$-loops can be controllably located in part of the substrate and
separated from the rest of the circuit based on ``0"-loops, i.e.
junctions where no additional $\pi$ phase shifts arise. This can be easily
made by depositing the MgO and CeO$_2$ seed layers on different parts of
the substrate, which will be also partly not covered by any seed layer.

As a test to show how the biepitaxial junctions could
be considered for preliminary tests  and  device implementation for quantum computing
without the topological restriction imposed by the
bicrystal technique, we refer to the structures
proposed in Ref.\cite{ioffe}as exemplary circuits.

The former is composed by a s-wave (S)- d-wave (D)- s-wave (S') double
junction connected with a capacitor and an ordinary ``0" Josephson
junction based on s-wave superconductors (the S-D'-S junction
generates the doubly degenerate state). The latter consists of a five
junction loop with a $\pi$ junction. Our technique would combine
the possibility of placing the ordinary ``0" junctions corresponding
to the MgO seed layer and to exploit the possible doubly degenerate state of
asymmetric 45$^\circ$ GB junctions  corresponding to the CeO$_2$ seed
layer to replace the S-D-S' system or  the $\pi$ junction
respectively. Our structure would be obviously composed only of HTS.
In Figs. 10a and 10b we show how  devices for instance such as those
proposed in Ref.\cite{ioffe} could
be  obtained by employing the biepitaxial technique respectively. The
application to the  five junction loop is straightforward (Fig.10b)
and the advantages of this structure have been
already discussed  in Ref. \cite{ioffe}$^b$. The biepitaxial technique can offer possible
alternatives for the realization of the structures above.
In particular  the double junctions of the original S-D-S'
system can be also replaced by a D'-D-D'' structure (Fig.10c) by exploiting
our technique, in contrast to the bicrystal technology which could
not give this possibility. Such a configuration could offer some
advantages, if we consider that asymmetric 45$^\circ$ bicrystal GB
Josephson junctions did not give systematic evidence of the
doubly degenerate state. The doubly
degenerate state seems to occur only in high quality low
transparency GB junctions \cite{ilic,tanaka} and it is known that S-I-D
junctions do not have double  periodicity of the critical current as a
function of the phase. \cite{tanaka}
A consequence  of a possible nodeless order parameter \cite{kirtley,bailey}
at the D-D' GB interface  could
be a closer similarity with a S-I-D junction with loss of the doubly degenerate
state. If this is the case, we speculate that the double junctions
structure for symmetry reasons would produce a leading term
in the Josephson coupling energy of the form E$_d$ cos2$\theta$
(double periodic) and that the possible dipolar component of the magnetic field
would be almost completely compensated in this configuration. \cite{ioffe}$^b$
This can be considered as an attempt to construct a ``microscopic" 2$\theta$-junction.
We finally notice that the topological advantages offered by
the biepitaxial junctions would therefore be crucial in both the
cases considered
for the realization of the structure in Fig. 10, and important
to reduce de-coherence effects. Bicrystal substrates would in fact
impose on the circuit additional junctions
required by the circuit design  and, as a consequence, generate
additional noise and de-coherence in the device.

\section{Conclusions}
The performance of the presented junctions and SQUIDs
demonstrates that significant improvements in the biepitaxial
technique are possible, and the resulting devices have potential
for applications. We have presented a phenomenology that is
consistent with the expected absence of $\pi$-loops in 45$^\circ$
[100] tilt and twist grain boundaries junctions. The use of a
CeO$_2$ rather than a MgO seed layer can produce $\pi$-loops in
the same  junction configurations. The versatility of the
biepitaxial technique has been recently used to obtain different
types of grain boundaries. The advantage of placing junctions in
arbitrary locations on the substrate without imposing any
restrictions on the geometry, and the ease of obtaining different
device configurations by suitably patterning the seed layer, make
the biepitaxial technique competitive for the testing of new
concept devices, such as those based on $\pi$-loops. Some simple
examples of situations in which $\pi$-loops can be suitably
produced in specific locations of a more complicated circuit have
also been discussed.

\acknowledgments This work has been partially supported by the
projects PRA-INFM ``HTS Devices" and SUD-INFM ``Analisi non
distruttive con correnti parassite tramite dispositivi
superconduttori" and by a MURST COFIN98 program (Italy). The
authors would like to thank Dr. E. Ilichev and A. Golubov for
interesting discussions on the topic.

\begin{figure}
\caption{ A schematic representation of the artificial grain
boundary structure. The boundary is obtained at the interface
between the [001] oriented YBCO film grown on the
 [110] MgO seed layer and the [103] YBCO film grown on the bare [110]  STO
substrate. In contrast with the 45$^\circ$ [001] tilt  bicrystal
junctions, in this case the order parameter orientations do not
produce an additional $\pi$ phase shift. }

\vspace{0.3in} \caption{ The CeO$_2$ seed layer produces an
artificial GB that can be seen as a result of two rotations: a
45$^\circ$ [100] tilt or twist followed by a 45$^\circ$ tilt
around the c-axis of the (001) film. For this junction
configuration a d-wave order parameter symmetry would produce
$\pi$ -loops. }

\vspace{0.3in} \caption{ Current vs voltage (I-V) characteristics
of the biepitaxial junction for temperature close to the critical
temperature. In the inset the I-V curve  at T = 4.2 K is shown. }

\vspace{0.3in} \caption{ a) Scheme of the seed layer
patterning,which allows the measuremment on the same chip of the
properties of a tilt junction and of junctions whose interface is
tilted in plane of an angle $\alpha$ = 30$^\circ$, 45$^\circ$ and
60$^\circ$ with respect the a- or b-axis of the (001) YBCO thin
film respectively. b) The I-V characteristics (measured at T =
4.2 K) of the microbridges reported in Fig. 4a. }

\vspace{0.3in} \caption{ Magnetic-field dependence of the
critical current of a [100] tilt biepitaxial dc-SQUID. The
absolute maximum is observed for zero field. A double-period
modulation is observed.The longer period modulation is the
diffraction pattern due to the magnetic field sensed by a single
junction, while the shorter period SQUID modulation is shown more
clearly in the inset (a). In the inset (b) I-V curves are shown
as a function of an externally  applied magnetic field at T = 4.2
K. A typical Fraunhofer-like dependence is evident. }

\vspace{0.3in} \caption{ Scanning SQUID microscope image of a
200x200$\mu$m$^2$ area along a grain boundary separating a (100)
region from a (103) region of a thin YBCO biepitaxial film grown.
The position of the grain boundary is indicated by the dashed
line. }

\vspace{0.3in} \caption{ Magnetic field dependence of the voltage
of a [100] tilt biepitaxial dc-SQUID at 77 K for different values
of the bias current. }

\vspace{0.3in} \caption{ Magnetic flux noise spectral densities
of a [100] tilt biepitaxial SQUID at T=77 K and T=4.2 K. The
SQUID, with an inductance L=13 pH, was modulated with a standard
flux-locked-loop electronics. The right axis shows the energy
resolution. Data at T = 4.2 K are compared with results on SQUIDs
based on [001] tilt biepitaxial junctions  from Ref. 25. }

\vspace{0.3in} \caption{ a) 3-dimensional view of a SQUID based
on 45$^\circ$ [100] tilt and twist GBs; no $\pi$-loops should
occur. b) Top view of $\pi$-SQUID based on 45$^\circ$ [001] tilt
GBs. c) 3-dimensional view of a $\pi$-SQUID based on GBs
resulting from two rotations: a 45$^\circ$ [100] tilt or twist
followed by a 45$^\circ$ [001] tilt }

\vspace{0.3in} \caption{ Scheme of the qubit structure proposed
in Ref.1 designed using  the biepitaxial grain boundaries
proposed in the paper. The double junctions of the original
S-D-S' system can be also replaced by D'-D-D''. }

\label{autonum}
\end{figure}

\narrowtext
\end{multicols}

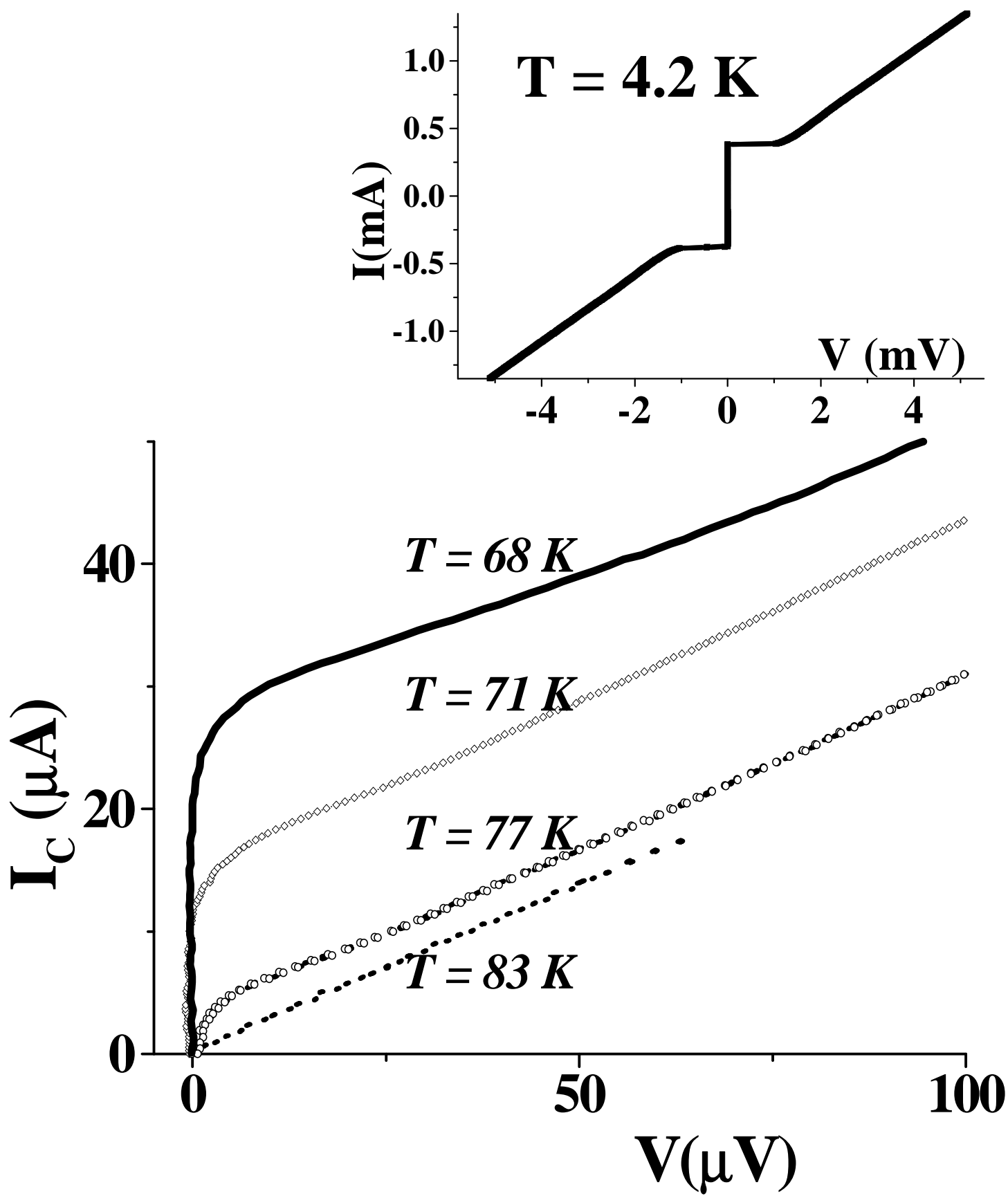
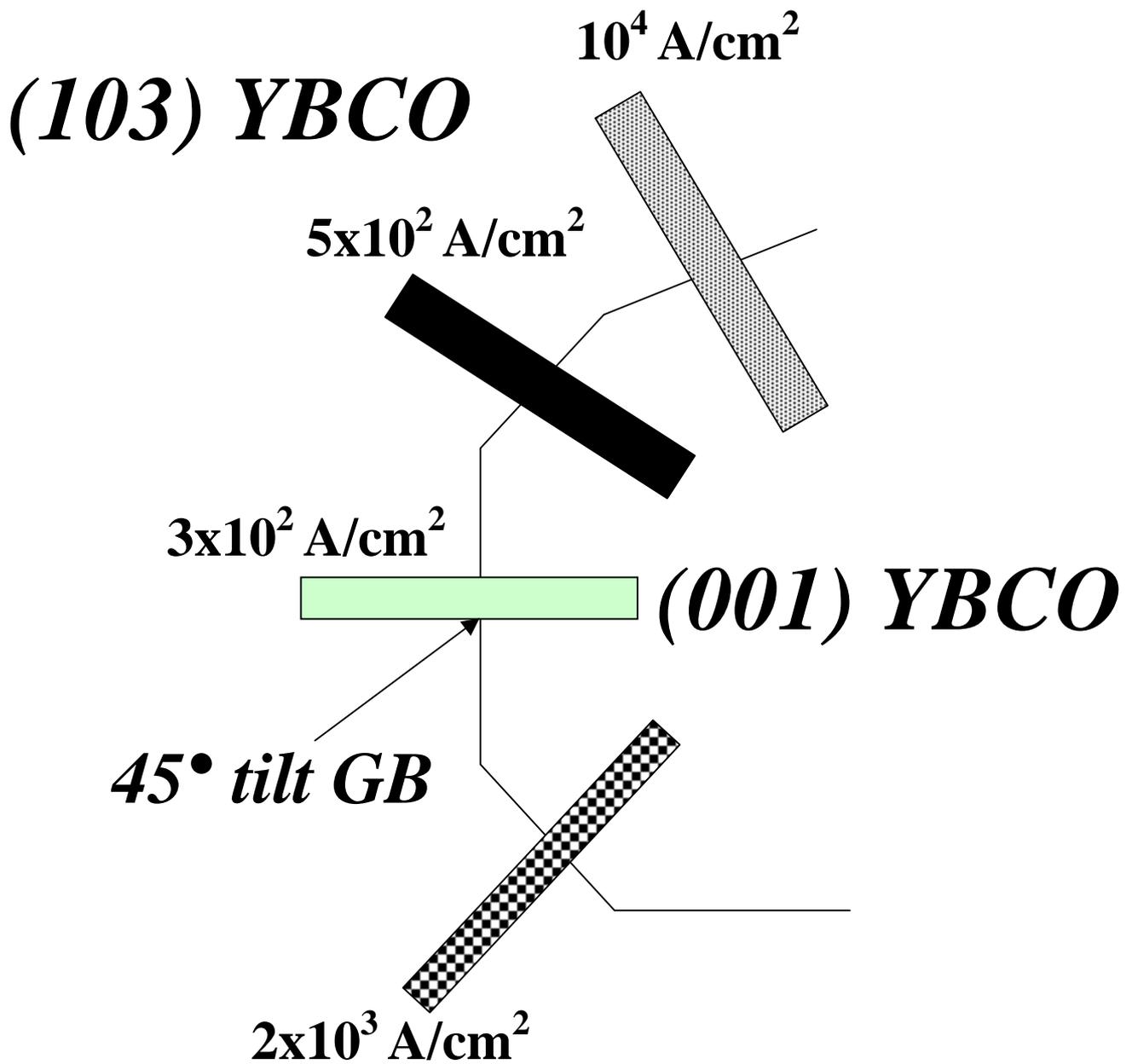
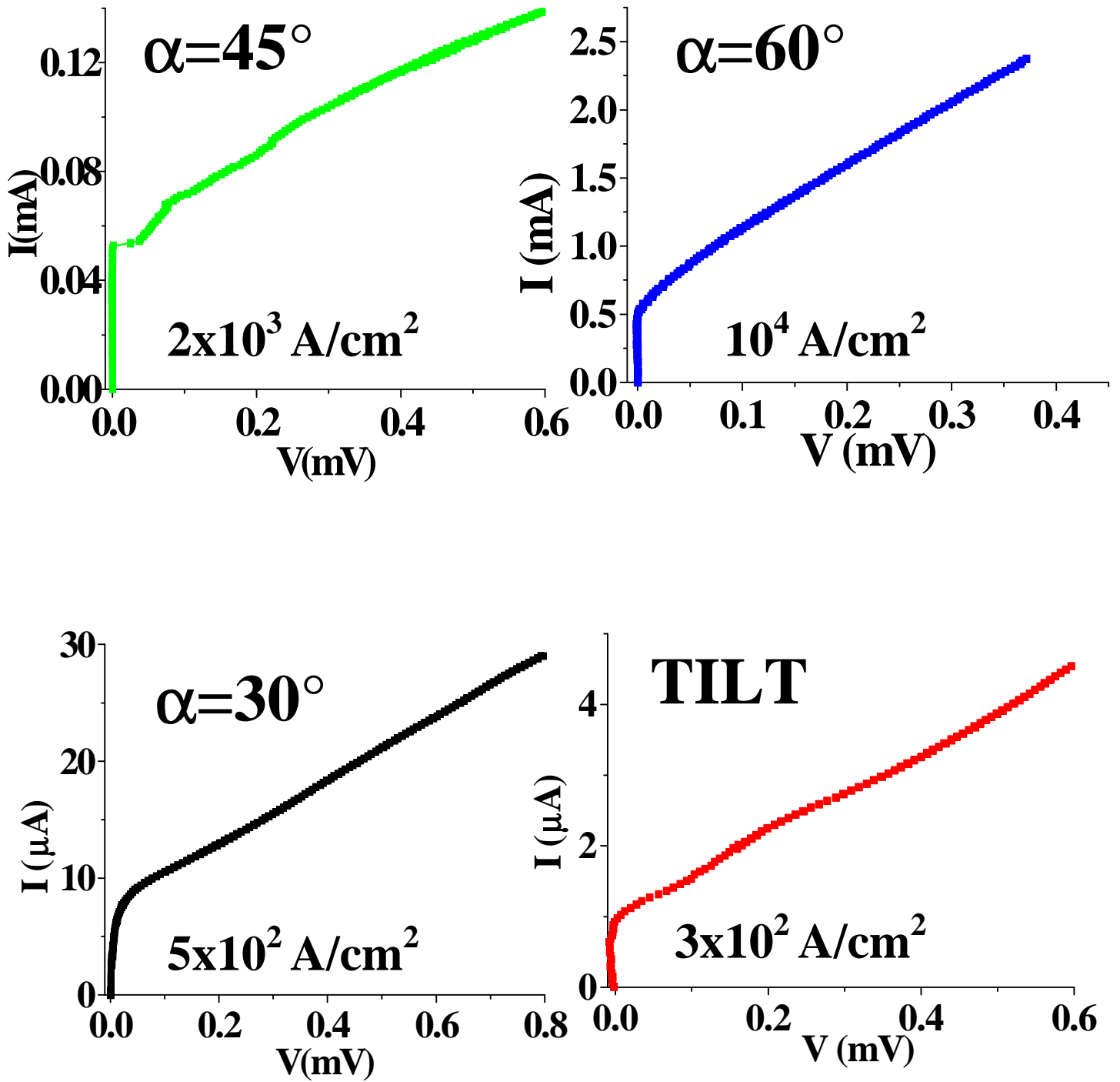
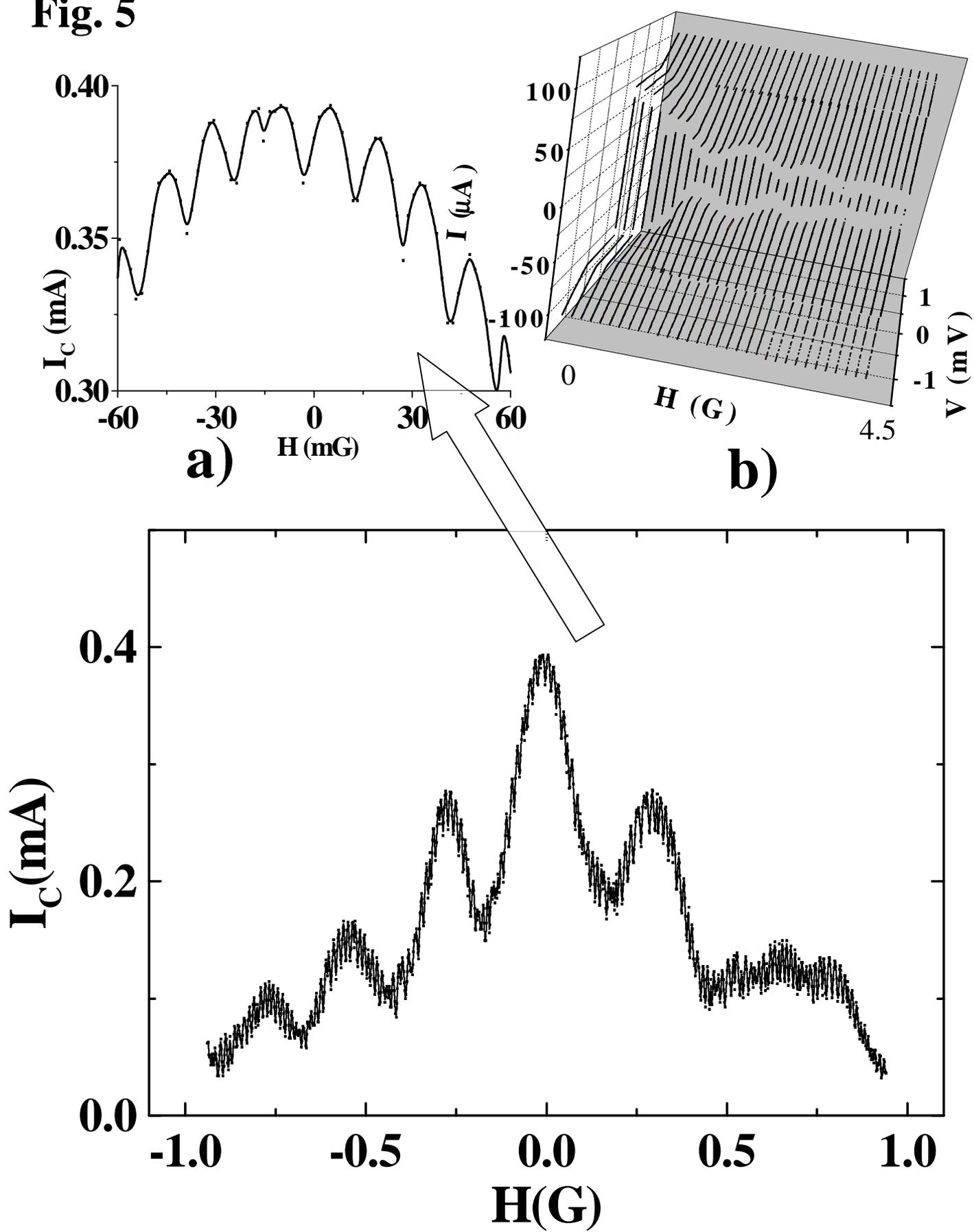
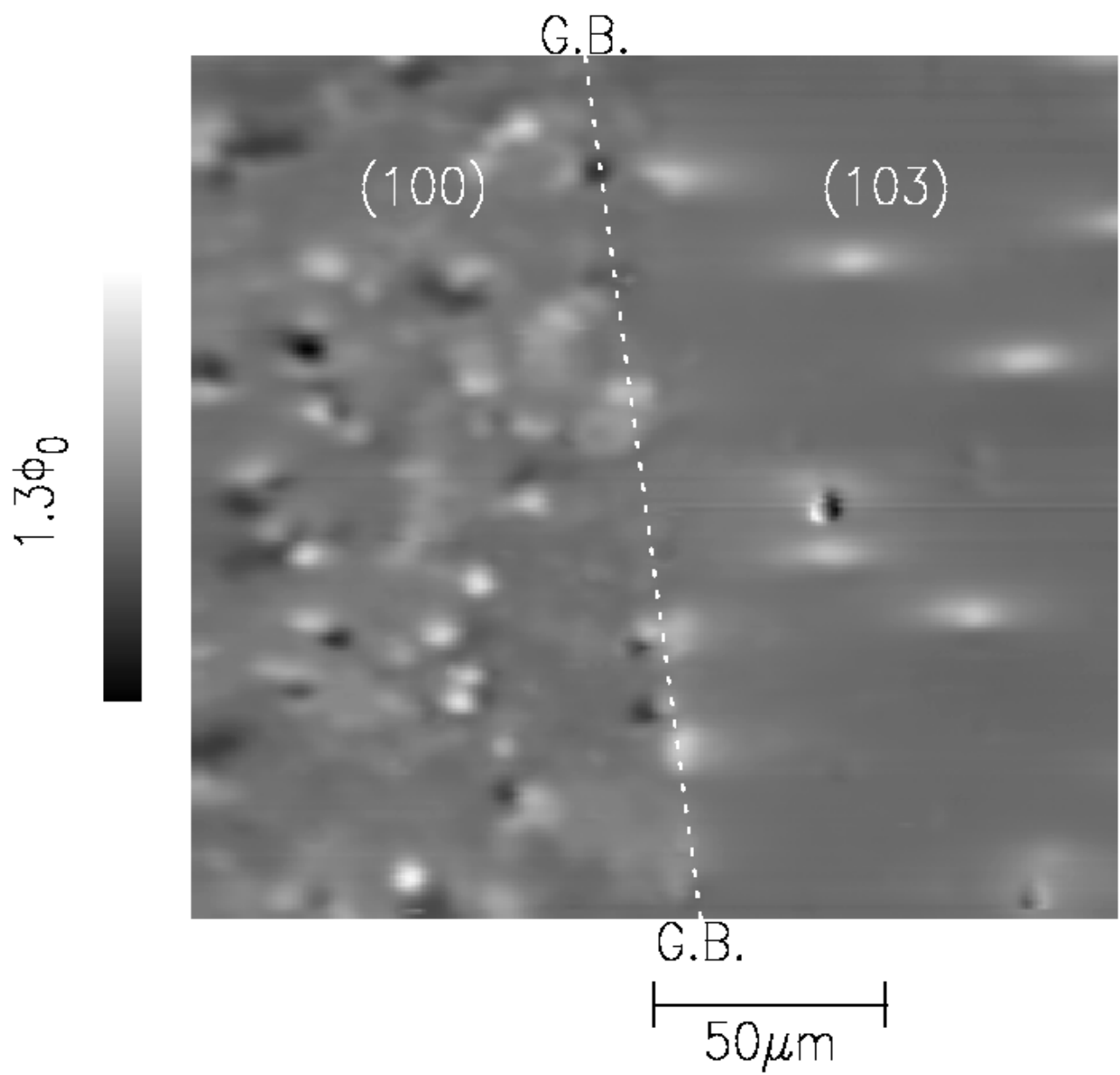
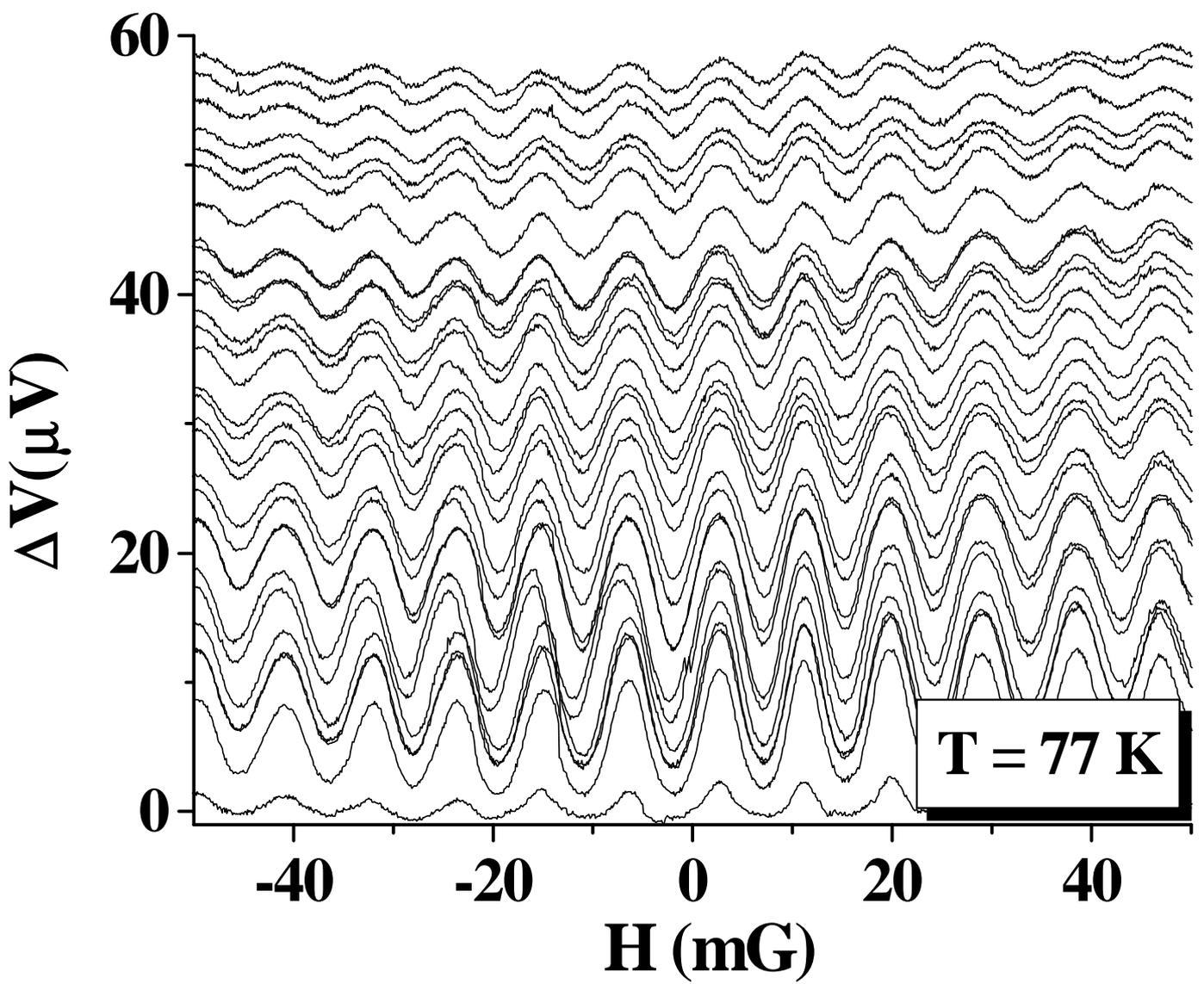
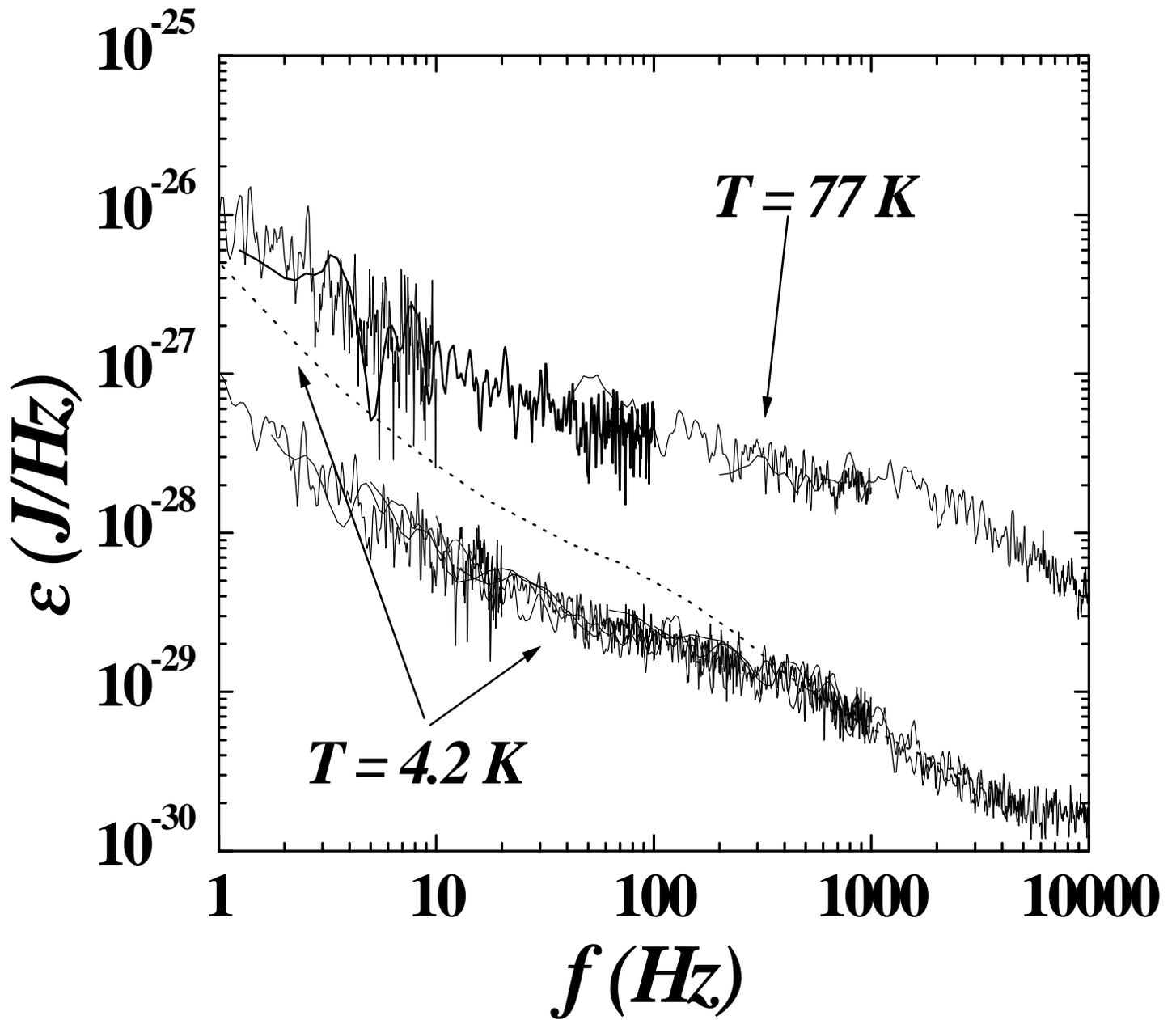
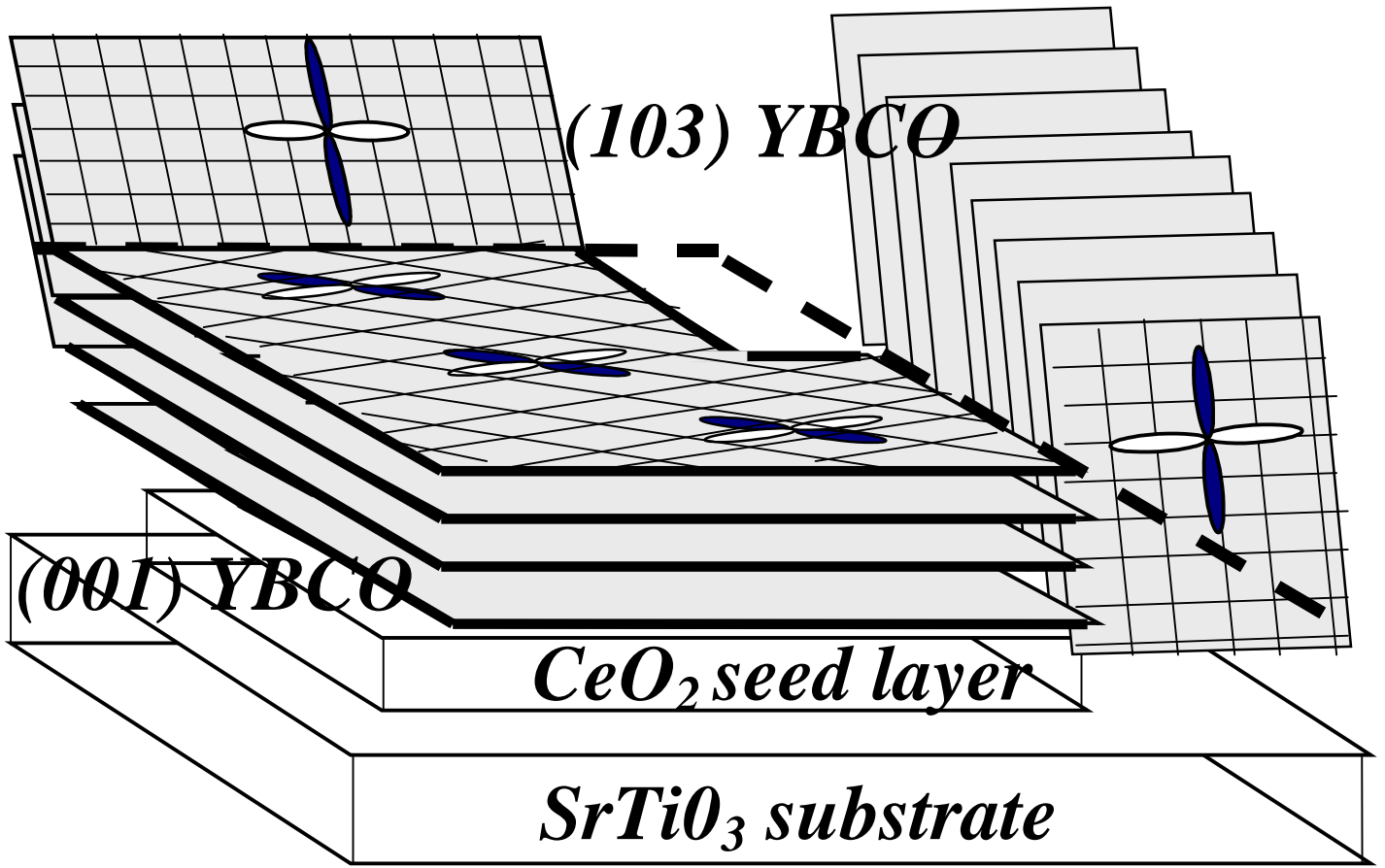
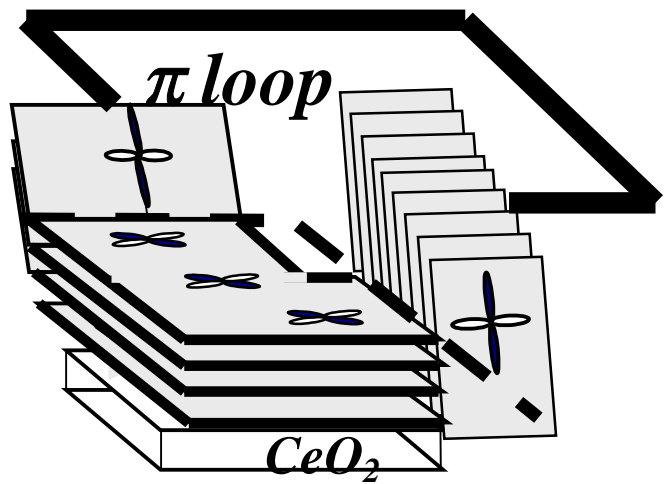
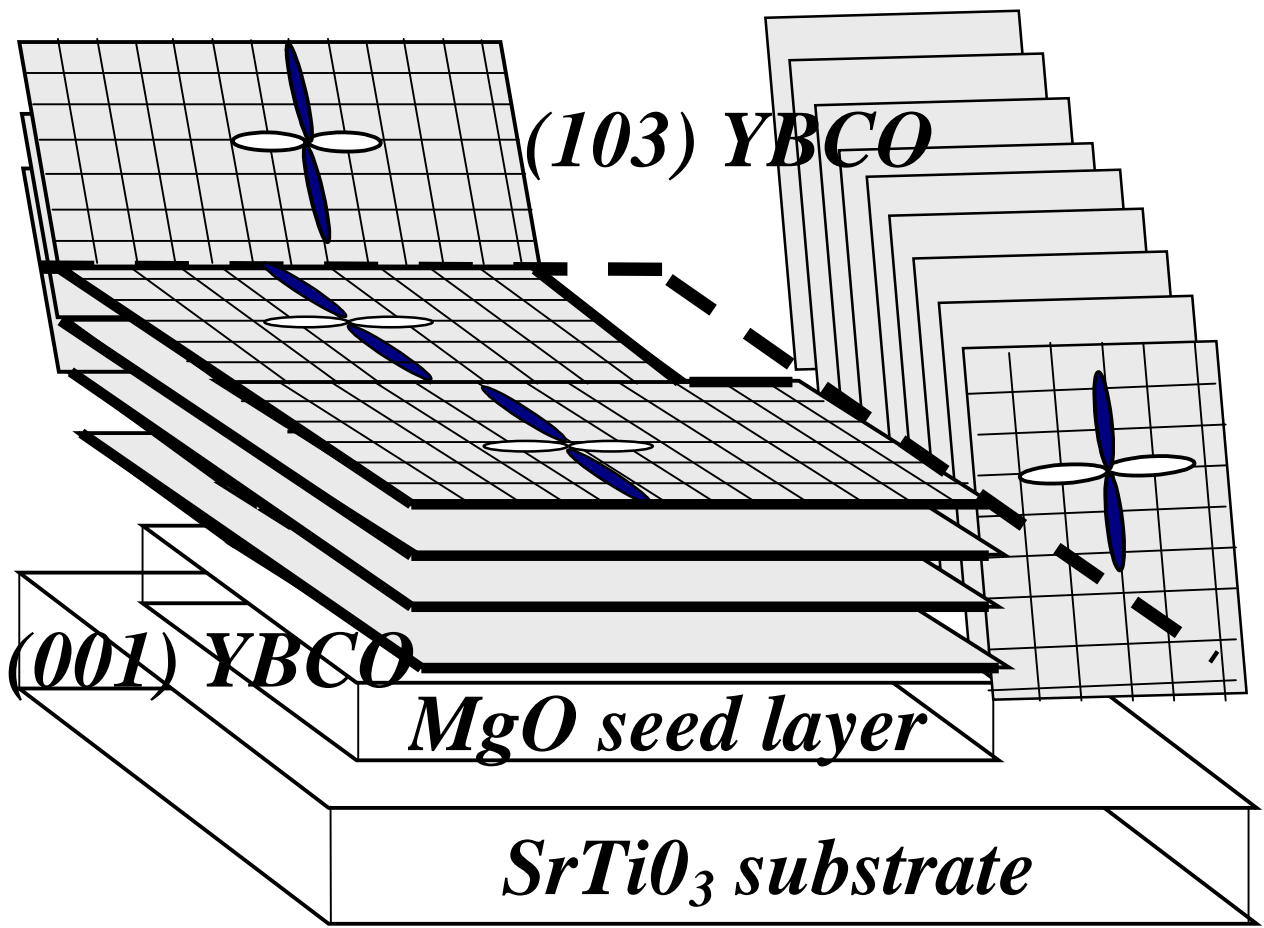
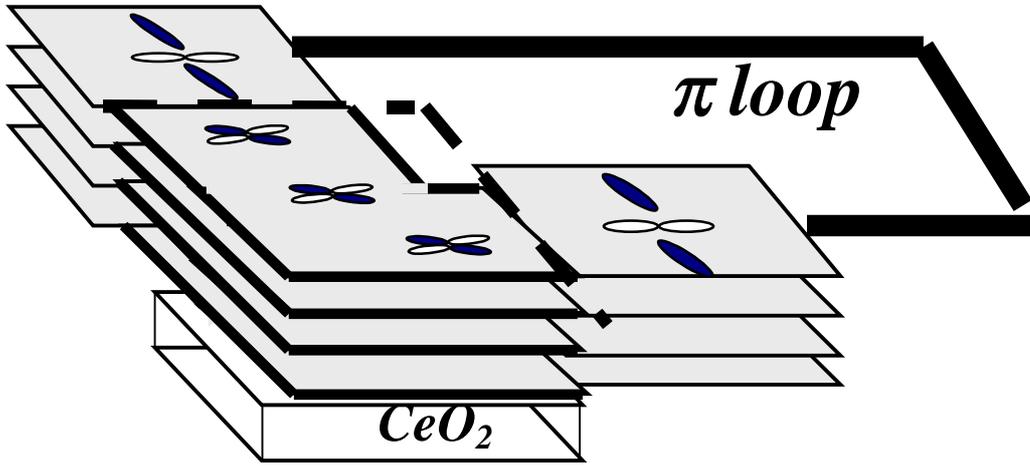
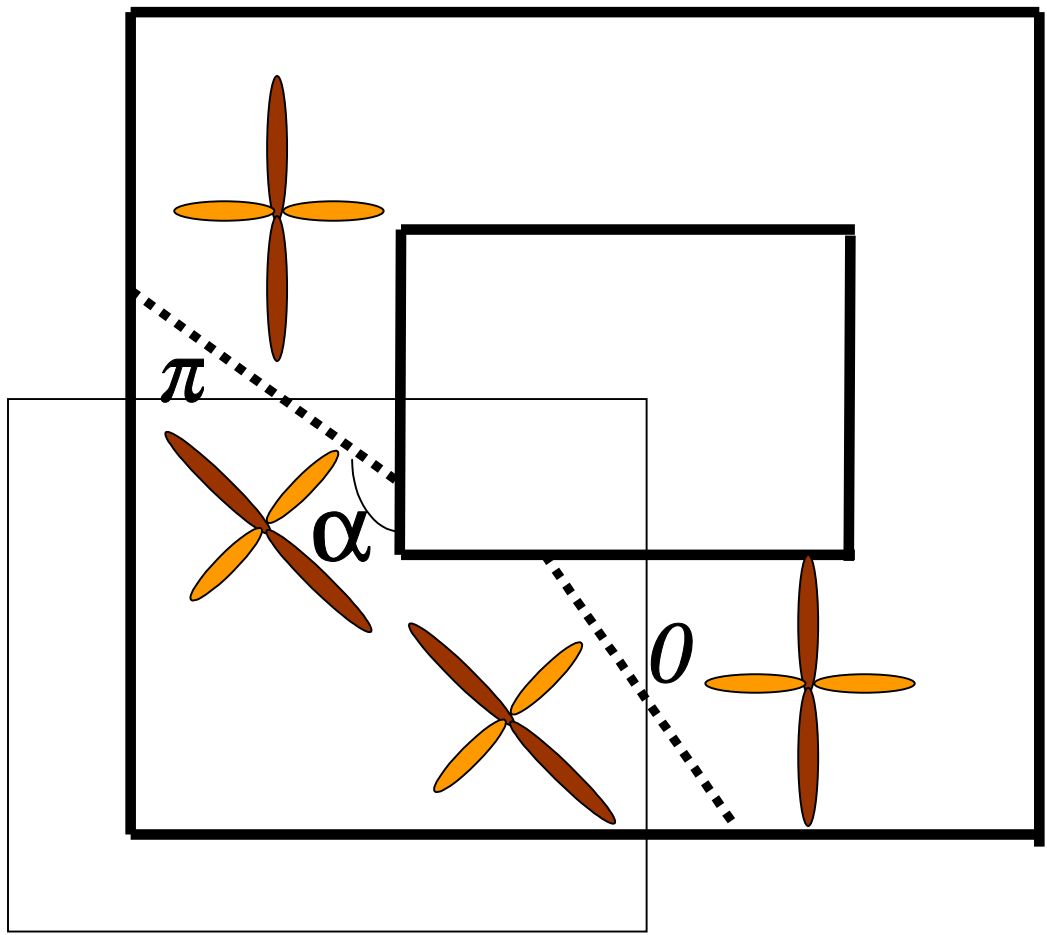
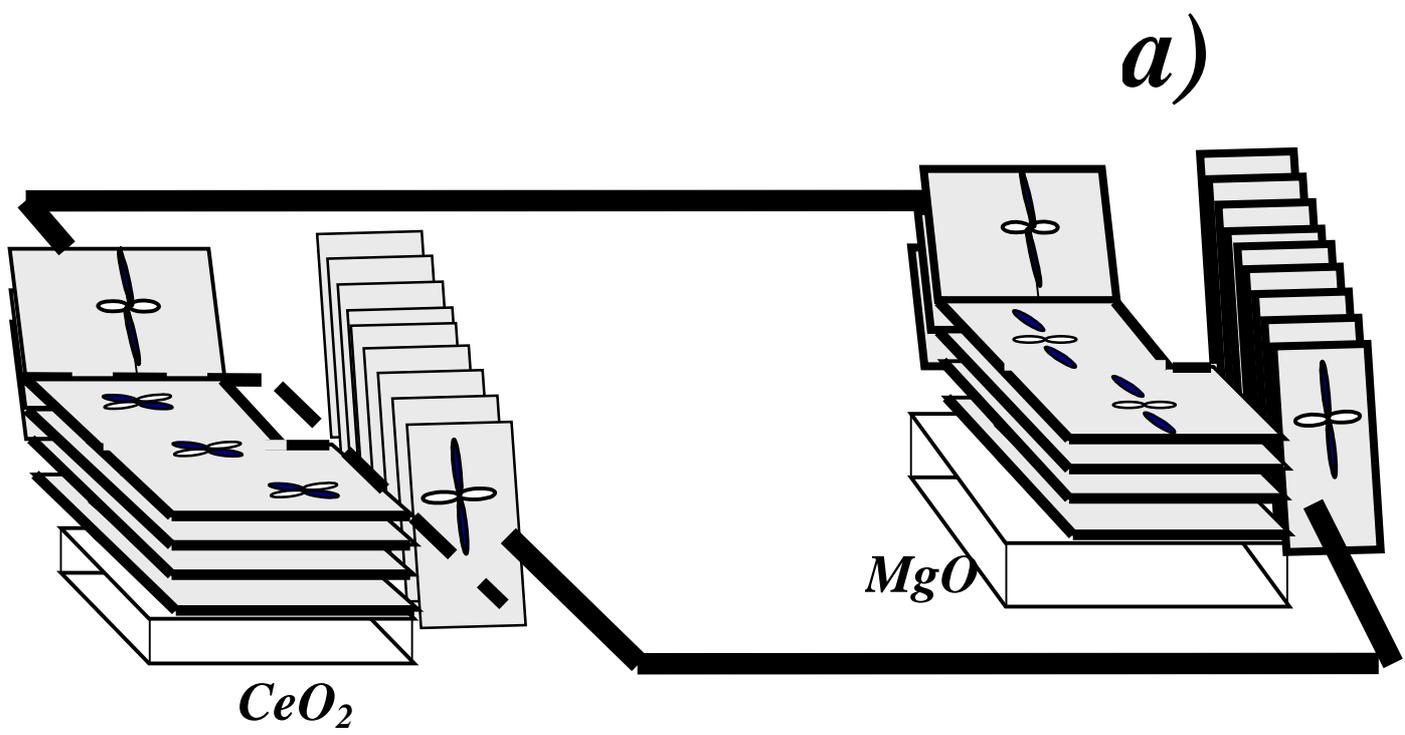
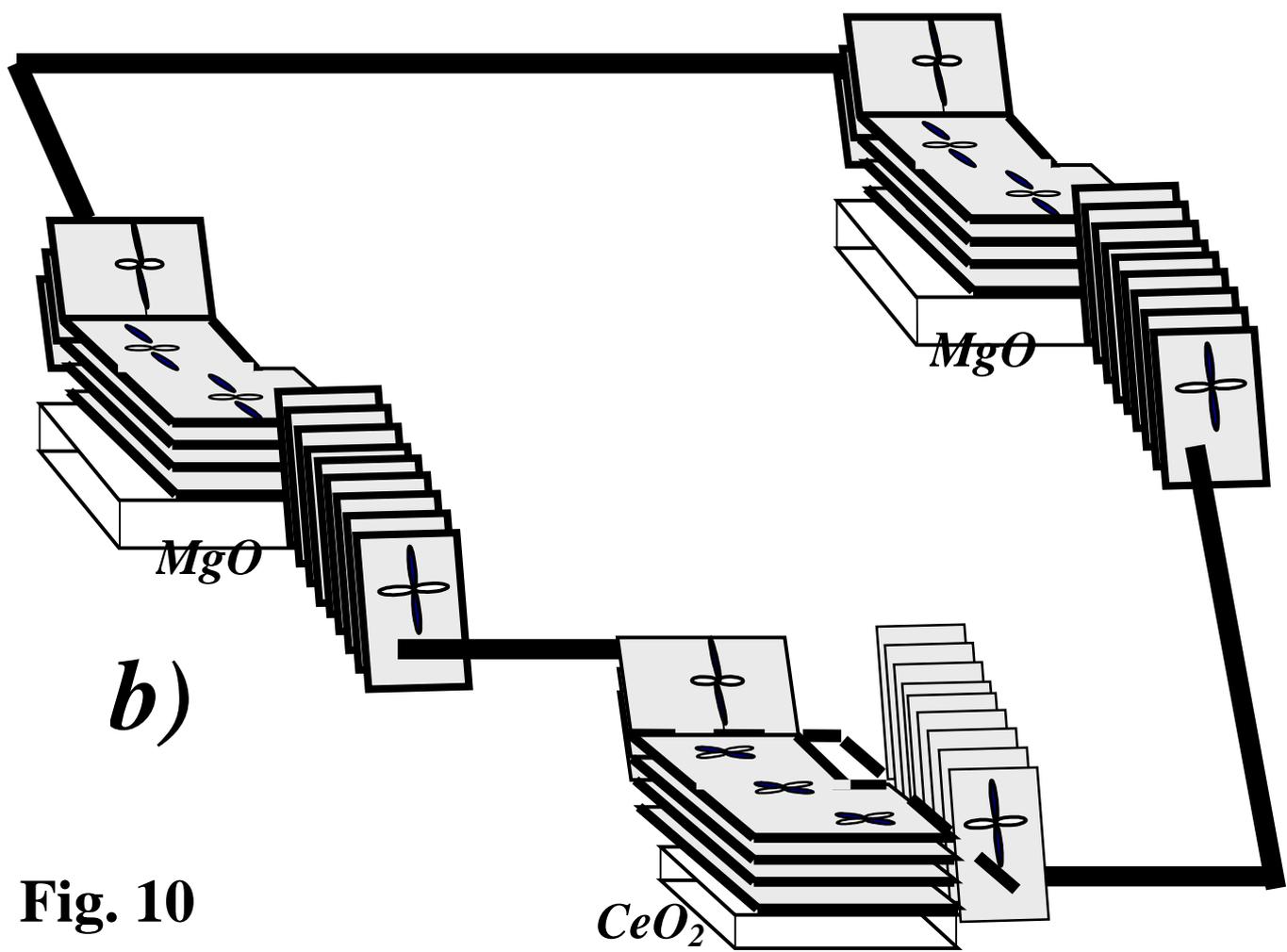
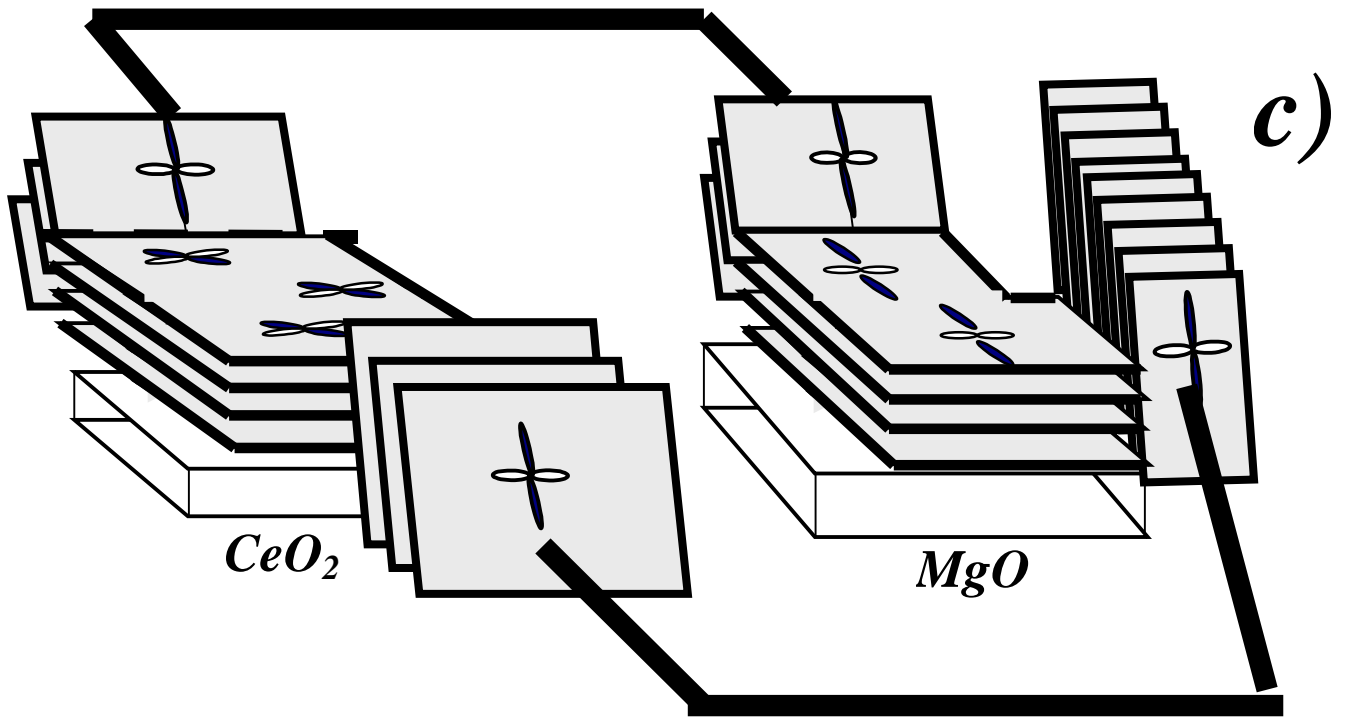
\end{document}